\documentclass[12pt]{article}
\usepackage{amsmath,amsfonts,amssymb,bbm}
\usepackage{tensor}
\usepackage[T1]{fontenc}
\usepackage[utf8]{inputenc}
\usepackage{lmodern}
\usepackage{authblk}
\usepackage[disable]{todonotes}
\usepackage{color}
\usepackage{multicol}
\usepackage{cite}
\usepackage{collref}
\usepackage[margin=3cm]{geometry}
\usepackage{hyperref}

\newcommand*\xbar[1]{%
  \hbox{%
    \vbox{%
      \hrule height 0.5pt % The actual bar
      \kern0.3ex%         % Distance between bar and symbol
      \hbox{%
        \kern-0.0em%      % Shortening on the left side
        \ensuremath{#1}%
        \kern-0.0em%      % Shortening on the right side
      }%
    }%
  }%
}

\newcommand\nn{\ensuremath{\mathcal N}}

\newcommand\email[1]{\thanks{\href{mailto:#1}{\nolinkurl{#1}}}}
\newcommand\td{\text{d}}

\newcommand\cA{{\cal A}}

\newcommand{\cJ}{{\cal J}}
\newcommand{\cL}{{\cal L}}

\newcommand{\bz}{\bar{z}}

\def\cH{\mathcal{H}}

\def\he{\hat{=}}
\def\bP{\bar{P}}
\def\bz{\bar{z}}
\def\bomega{\bar{\omega}}
\def\bL{\bar{L}}

\def\bY{\bar{Y}}

\def\cY{\mathcal{Y}}
\def\cA{\mathcal{A}}
\def\cQ{\mathcal{Q}}
\allowdisplaybreaks[1]

\newcommand{\p}{\partial}

\newcommand{\be}{\begin{equation}}
\newcommand{\ee}{\end{equation}}
\newcommand{\bea}{\begin{eqnarray}}
\newcommand{\eea}{\end{eqnarray}}
\def\nn{\nonumber}
\newcommand{\half}{\frac{1}{2}}


\author[a]{Pujian Mao\email{pujian.mao@ulb.ac.be}\,}
\author[b]{Xiaoning Wu\email{wuxn@amss.ac.cn}\,}
\author[c,d]{Hongbao Zhang\email{hzhang@vub.ac.be}\,}

\affil[a]{\,Service de Physique Th\'eorique et Math\'ematique, Universit\'e Libre de Bruxelles and International Solvay Institutes, Campus de la Plaine, CP 231, B-1050 Bruxelles, Belgium}
\affil[b]{\,Institute of Mathematics, Academy of Mathematics and System Science, Chinese Academy of Sciences, Beijing 100190, China}
\affil[c]{\,Department of Physics, Beijing Normal University, Beijing 100875, China}
\affil[d]{\,Theoretische Natuurkunde, Vrije Universiteit Brussel, and The International Solvay Institutes, Pleinlaan 2, B-1050 Brussels, Belgium}
\title{\bf Soft hairs on isolated horizon implanted by electromagnetic fields\\}
\date{}

\begin{document}
 \maketitle
 \thispagestyle{empty}

\begin{abstract}
Inspired by the recent proposal of soft hair on black holes in Phys. Rev. Lett. 116, 231301 (2016), we have shown that an isolated horizon carries soft hairs implanted by electromagnetic fields. The solution space and the asymptotic symmetries of Einstein-Maxwell theory have been worked out explicitly near isolated horizon. The conserved current has been computed and an infinite number of near horizon charges have been introduced from the electromagnetic fields associated to the asymptotic $U(1)$ symmetry near the horizon, which indicates the fact that isolated horizon carries a large amount of soft electric hairs. The soft electric hairs, \textit{i.e.} asymptotic $U(1)$ charges, are shown to be equivalent to the electric multipole moments of isolated horizons. It is further argued that the isolated horizon supertranslation is from the ambiguity of its foliation and an analogue of memory effect on horizon can be expected.
\end{abstract}

\cleardoublepage

\section{Introduction}

\label{sec:introduction}

The Bondi-Metzner-Sachs(BMS) group\cite{Bondi:1962px,Sachs:1962wk,Sachs:1962zza} has been discovered for more than fifty years. Though enormous efforts \cite{Newman:1966ub,Mccarthy:1972ry,Crampin:1974aw,Girardello:1974sq,Piard:1977vh,Barnich:2009se,Barnich:2010eb,Barnich:2011mi,Barnich:2016lyg,Compere:2016jwb} have been given to uncover this mysterious and fascinating asymptotic symmetry group in the past half century, it is still not, we believe, well understood especially for the quantum aspect of this group. On the one hand, we need a special treatment for the field theory with boundaries \cite{Gervais:1976ec}. On the other hand, as an ``improper'' gauge transformation \cite{Regge:1974zd,Benguria:1976in} on a manifold with a boundary, physical observables need not be invariant under the BMS transformation. How to count this new set of dynamical degrees of freedom is a quite difficult question. Moreover, the enhanced boundary degrees of freedom are always accompanied with the gravitational propagating degrees of freedom. To be more precise, the asymptotic symmetry of 4 dimensional asymptotically flat spacetime will be reduced to Poincar\'{e} group \cite{Ashtekar:1978zz} at spatial infinity who has no dynamics \cite{Geroch:1977jn}, unlike the 3 dimensional asymptotically anti de Sitter case \cite{Brown:1986nw} which has a better understanding from the holographic viewpoint \cite{Strominger:1997eq}.

Recently, a new scenario has been proposed by Strominger et al \cite{Strominger:2013jfa,He:2014laa}. They argued that gravity S-Matrix should have BMS symmetry and found a deep connection between BMS supertranslation and Weinberg's soft graviton theorem \cite{Weinberg:1965nx}. Such a treatment has been extended and succeeded in other gauge theories shortly \cite{He:2014cra,Strominger:2013lka,Lysov:2014csa,Kapec:2014zla,Campiglia:2015qka,Kapec:2015ena,Strominger:2015bla,He:2015zea,Dumitrescu:2015fej,Campiglia:2016hvg}, which gives  the asymptotic symmetry a new lease of life. Apart from the fruitful result, this scenario is highly sensitive to the existence of black hole in the bulk. On the symmetry level, the asymptotic symmetry group on future null infinity(BMS$^+$) and the asymptotic symmetry group on past null infinity(BMS$^-$) are isomorphic to each other but they are independent symmetries on different null infinities. To be a symmetry of the S-matrix, a canonical relation between BMS$^+$ and BMS$^-$ is required to act on both incoming and outgoing state. Only in a finite neighborhood of the Minkowski spacetime, a canonical identification between elements of BMS$^+$ and BMS$^-$ can be achieved \cite{Christodoulou:1993uv}. The presence of black hole, though the spacetime is still asymptotically flat, will definitely challenge this identification. At the quantum level, the unitarity is not guaranteed due to black hole formation, which is known as information paradox \cite{Hawking:1976ra}.

Nevertheless, the obstacle of black hole formation does not block the progress of understanding the asymptotic symmetries, but sheds new insights to black hole physics \cite{Hawking:2015qqa,Hawking:2016msc,Donnay:2015abr,Eling:2016xlx,Averin:2016ybl,Hooft:2016itl,Compere:2016hzt,Kapec:2016aqd,Hotta:2016qtv,Hawking:2016sgy,Sheikh-Jabbari:2016unm,Blau:2016juv,Afshar:2016uax,Grumiller:2016kcp,Cardoso:2016ryw,Donnay:2016ejv,Sheikh-Jabbari:2016npa,Carlip:2016lnw,Cai:2016idg,Strominger:2016wns,Shi:2016jtn,Afshar:2016kjj,Bergshoeff:2016soe,Hajian:2016iyp}. As a null hypersurface, the black hole horizon can be treated as an inner boundary. The symmetries near horizon create new dynamical degrees of freedom. However, the well-known uniqueness theorem \cite{Chrusciel:2012jk} tells us that the non-zero conserved charge of a stationary black hole in Einstein-Maxwell theory should be only $M$, $J$ and $Q$. A stationary horizon seems insensitive to the new dynamical degrees of freedom. As also pointed in \cite{Hawking:2016msc}, the dynamical process such as black hole formation/evaporation will cause the changing of vacuum state of gravity. This means that one needs to investigate non-stationary process instead of just considering stationary black hole. In order to do so, the isolated horizon should be a better candidate for understanding the near horizon physics. An isolated horizon was introduced in \cite{Ashtekar:1998sp}, who describes later state of black holes. All dynamical process in the neighborhood of the horizon has been almost settled down, but the space-time far away from the horizon can be still dynamical. Consequently, the isolated horizon framework serves as a more realistic resolution of black hole physics and is playing an important role in numerical simulations. Though the exact symmetry of an isolated horizon is not rich enough \cite{Ashtekar:2001is}, it was shown in \cite{Ashtekar:2004gp} that the isolated horizon structure has some ambiguity which is supposed to count for the recently revealed near horizon supertranslation in \cite{Donnay:2015abr}. The action of a supertranslation on the isolated horizon can map one set of isolated horizon structure into another one which is equivalent to a dynamical process \textit{e.g.} radiation crossing the horizon. This can be understood as a isolated horizon analogue of memory effect at null infinity \cite{Strominger:2014pwa}. Hence, it would be highly meaningful to have a fully asymptotic analysis of an isolated horizon.

In this paper, the asymptotic structure of Einstein-Maxwell system near the isolated horizon will be work out explicitly. We will show a tractable solution space and asymptotic symmetries of Einstein-Maxwell theory near the isolated horizon $\cH$. The asymptotic symmetries consist of supertranslation, superrotation, and the asymptotic $U(1)$ symmetry which is an expected enhancement of the new discovery in \cite{Donnay:2015abr}. Those symmetries form a closed algebra. The local conserved current associated with the asymptotic symmetries will be constructed. To have a concrete physical interpretation, we would restrain ourself in a special class of the solution space which is the approximation of a Schwarzschild black hole surrounded by electromagnetic fields. In such a physical specification, one can define an infinite number of conserved charges respecting to the asymptotic $U(1)$ symmetry. According to the terminology of \cite{Hawking:2016msc}, apart from the electric charge, we would call them as soft electric hairs. It's further shown that those soft hairs are equivalent to the electric multipole moments of isolated horizons defined in \cite{Ashtekar:2004gp}. The supertranslation charges can be also introduced but they are vanishing except the zero mode. Remarkably, the presence of Maxwell fields does not affect the zero mode of the supertranslation charge. This is direct evidence that the soft electric hairs are implanted by soft photons who contain no energy. The existence of soft hairs reveals new dynamical degrees of freedom meaning that isolated horizons with different soft electric hairs should be considered as different physical states. This may inspire a direct counting of horizon degrees of freedom.

This paper is organized as follows. In next section, after a review of isolated horizon, the solution space of Einstein-Maxwell theory near an isolated horizon will be worked out explicitly. Section 3 presents the asymptotic symmetry structure of Einstein-Maxwell theory and the conserved current associated with those asymptotic symmetries. The influence of the presence of electromagnetic fields around Schwarzschild black hole, as a special physical approximation, will be illustrated in section 4. The near horizon supertranslation and the ambiguity of an isolated horizon will be discussed in section 5 to depict an analogue of memory effect on horizon. The last section will be devoted to some discussions. There is also an appendix listing the details of the solution space of Einstein-Maxwell theory near an isolated horizon.

\section{Solution space of Einstein-Maxwell theory near isolated horizon}\label{soluchapter}
The concept of isolated horizons was introduced to approximate event horizons of black holes at late stages of gravitational
collapse and of black hole mergers when back-scattered radiation falling into the black hole can be neglected \cite{Ashtekar:1998sp}.

In the present work, we will use the Newman-Penrose(NP) formalism \cite{Newman:1961qr} which is a tetrad system with four null basis $l,n,m,\xbar m$ satisfying orthogonality conditions $l\cdot m=l\cdot \xbar m=n\cdot m=n\cdot \xbar m=0$ and normalization conditions $l\cdot n=-m\cdot\xbar m=1$.
The various Ricci rotation-coefficients in tetrad system, now called the spin coefficients, are designated by a list of special symbols as follows:
\bea\label{spinconnection}
&&\kappa=l^\nu m^\mu\nabla_\nu l_\mu,\;\;\pi=-l^\nu \bar{m}^\mu\nabla_\nu n_\mu,\;\;\epsilon=\half(l^\nu n^\mu\nabla_\nu l_\mu - l^\nu \bar{m}^\mu\nabla_\nu m_\mu),\nn\\
&&\tau=n^\nu m^\mu\nabla_\nu l_\mu,\;\;\nu=-n^\nu \bar{m}^\mu\nabla_\nu n_\mu,\;\;\gamma=\half(n^\nu n^\mu\nabla_\nu l_\mu - n^\nu \bar{m}^\mu\nabla_\nu m_\mu),\nn\\
&&\sigma=m^\nu m^\mu\nabla_\nu l_\mu,\;\;\mu=-m^\nu \bar{m}^\mu\nabla_\nu n_\mu,\;\;\beta=\half(m^\nu n^\mu\nabla_\nu l_\mu - m^\nu \bar{m}^\mu\nabla_\nu m_\mu),\nn\\
&&\rho=\bar{m}^\nu m^\mu\nabla_\nu l_\mu,\;\;\lambda=-\bar{m}^\nu \bar{m}^\mu\nabla_\nu n_\mu,\;\;\alpha=\half(\bar{m}^\nu n^\mu\nabla_\nu l_\mu - \bar{m}^\nu \bar{m}^\mu\nabla_\nu m_\mu),\nn
\eea
where $\nabla_\mu$ is the covariant derivative. It would be helpful to summarize briefly the geometric meaning of the spin coefficients introduced in \cite{Newman:1961qr,Chandrasekhar}.

Consider a null congruence generated by $l_\mu$, the spin coefficient $\kappa$ is related to the first curvature by
\be
l^\nu\nabla_\nu l_\mu=-\kappa \xbar m_\mu -\xbar\kappa m_\mu+(\epsilon+\xbar\epsilon)l_\mu.
\ee
Thus $\kappa=0$ yields a null geodesic congruence. In the case of an affine parameter, namely $\epsilon=0$ which can be achieved by a rescaling of $l_\mu$, we have
\be
\rho+\xbar\rho=-\nabla_\mu l^\mu,\;\;(\rho-\xbar\rho)^2=-\nabla^\mu l^\nu(\nabla_\mu l_\nu - \nabla_\nu l_\mu),\ee
\be(\rho+\xbar\rho)^2+4\sigma\xbar\sigma=\nabla^\mu l^\nu(\nabla_\mu l_\nu + \nabla_\nu l_\mu).
\ee
If one follows the geodesic congruence of $l_\mu$, the expansion, the rotation and the shear are measured by Re$(\rho)$, Im$(\rho)$ and $\sigma$ respectively. Moreover, $\tau$ describes the change of the direction $l_\mu$ along the direction $n_\nu$ following the relation
\be
n^\nu \nabla_\nu l_\mu=-\tau \xbar m_\mu - \xbar\tau m_\mu + (\gamma+\xbar\gamma)l_\mu.
\ee
The spin coefficient $\nu$, $\mu$, $\lambda$, $\pi$ are the analogue of $\kappa$, $-\rho$, $-\sigma$, $\tau$ when the geodesic congruence generator is $n_\mu$ instead of $l_\mu$.

In NP formalism the Weyl tensor and the field-strength \textit{i.e.} the physical fields in Einstein-Maxwell system, are organized as follows:
\bea\label{weyl}
&&\Psi_0=-C_{abcd}l^am^bl^cm^d,\;\;\Psi_1=-C_{abcd}l^an^bl^cm^d,\;\;\Psi_2=-C_{abcd}l^am^b\xbar m^cn^d,\nn\\
&&\Psi_3=-C_{abcd}l^an^b\xbar m^cn^d,\;\;\Psi_4=-C_{abcd}n^am^bn^c\xbar m^d,\nn\\
&&\phi_0=F_{ab}l^am^b,\;\;\phi_1=\half F_{ab}(l^an^b+\xbar m^am^b),\;\;\phi_2=F_{ab}\xbar m^an^b.\nn
\eea
The ten components of the Ricci tensor are defined in terms of four real and three complex scalars:
\be
\Phi_{00}=-\half R_{11},\;\;\Phi_{11}=-\dfrac{1}{4}(R_{12}+R_{34}),\;\;\Phi_{22}=-\half R_{22},\;\;\Lambda=\dfrac{1}{24}R=\dfrac{1}{12} (R_{12}-R_{34}),\ee
\be\begin{split}
&\Phi_{02}=-\half R_{33},\;\;\Phi_{01}=-\half R_{13},\;\;\Phi_{12}=-\half R_{23},\\
&\Phi_{20}=-\half R_{44},\;\;\Phi_{10}=-\half R_{14},\;\;\Phi_{21}=-\half R_{24},
\end{split}\ee
where $\Lambda$ is the cosmological constant and $\Phi_{ab}$ is the adapted stress-energy tensor. In Einstein-Maxwell theory $\Phi_{ab}=\phi_a\xbar\phi_b$.
The equations governing those quantities are called Newman-Penrose equations which are the analogue of Cartan's structure equations in the standard tetrad system.
We will adopt the convention of \cite{Chandrasekhar} throughout this paper and will directly refer to the notations and equations therein. Let us choose $n$ to be the normal vector\footnote{Our convention of the normal vector is different from \cite{Ashtekar:2001is,Ashtekar:2001jb} to be consistent with the notation at null infinity. To compare with \cite{Ashtekar:2001is,Ashtekar:2001jb}, one just needs to exchange $l$ and $n$.}  of the isolated horizon $\cH$. According to \cite{Ashtekar:2001is,Ashtekar:2001jb}, the definition of a generic Isolated Horizon in Newman-Penrose formalism is
\vspace{0.5cm}

\textbf{Definition 1}: \textit{~A 3-dimensional null sub-manifold $\mathcal{H}$ is called an isolated horizon if
            \newline(1)~ $\mathcal{H}$ is diffeomorphic to the product $\mathbb{S}\times R$, where $\mathbb{S}$ is a $2$-dimensional space-like manifold, and the fibers of the projection
            \begin{equation*}
			\Pi : \mathbb{S} \times R \rightarrow \mathbb{S}
			\end{equation*}
			are null curves in $\mathcal{H}$;
			\newline(2)~ the expansion of its normal vector $n$ vanishes everywhere on $\mathcal{H}$;
			\newline(3)~ Einstein's equations hold on $\cH$ and the stress-tensor satisfies $\Phi_{22}\he\Phi_{12}\he\Phi_{21}\he0$, where $\hat{=}$ means on horizon $\mathcal{H}$ only;
			\newline(4)~ the entire geometry of $\mathcal{H}$ and the gauge potential $A_\mu$ are stationary, \textit{i.e.} time-independent.}
\vspace{0.5cm}

To admit an isolated horizon structure, the horizon data will have several constraints. Since $\mathcal{H}$ is null, the normal vector $n$ is the generator of null geodesic on the horizon $\cH$. Hence, the spin coefficients relation
\bea
n^\mu\nabla_\mu n=-(\gamma+\xbar\gamma)n+\nu m+\xbar\nu\xbar m\nn
\eea
gives $\nu\he0$. $k_{(n)}:=-(\gamma+\xbar\gamma)$ will be called surface gravity of $\mathcal{H}$. The orthogonality and normalization conditions of the tetrad system now uniquely specify a two-parameter subset of geodesic of a null geodesic congruence by displacement vectors $m,\xbar m$. From the definition of the spin coefficients, requirement (2) implies $\mu\he0$; (3) fixes the complex scalar $\phi_2\he0$; the last condition requires the Lie derivative of all spin coefficients and gauge fields alone $n$ vanish on the horizon. Moreover, $k_{(n)}$ is a constant on the horizon. When $k_{(n)}=0$, it will be called an extremal isolated horizon. Those constraints are derived explicitly in \cite{Ashtekar:2001jb,Ashtekar:2001is}.

In a Bondi-like coordinate ($u,r,z,\bz$), the horizon will be located at $r=0$. We will further choose the following gauge and boundary conditions on Maxwell potential $A_\mu$ as $A_r=0,A_u\he0$. The full solution of Newman-Penrose equations near a non-extremal isolated horizon has been derived in Appendix \ref{solution}. We summarize as following:

\vspace{0.5cm}

\textit{A solution of Einstein-Maxwell theory in the neighborhood of an isolated horizon will be specified by the initial data $\Psi_0(r,z,\bz),\;\phi_0(r,z,\bz)$ and the data on $\mathbb{S}$ including its metric $q_{z\bz}=\frac{1}{P(z,\bz)\bP(z,\bz)}$, its extrinsic curvature $\tau(z,\bz)$, electromagnetic fields $A_u(z,\bz),\;A_z(z,\bz)$, together with the constant surface gravity $k_{(n)}$ \footnote{The local existence for the characteristic initial value has been proven in \cite{Luk:2011vf} in a neighborhood in the future of the light cone for vacuum Einstein equations. Using similar method, it is not difficult to extend the local existence to Einstein-Maxwell equations with the conditions we have set on Maxwell fields.}.}

\vspace{0.5cm}

The near horizon metric in ($+,-,-,-$) signature and gauge fields are given by
\bea
&&ds^2=[4\gamma_0 r + (\Psi_2^0 + \xbar\Psi_2^0 +2 \phi_1^0\xbar\phi_1^0)r^2 + O(r^3)]du^2 + 2dudr \nn\\
&&\hspace{1cm}+ [4\frac{\xbar\tau_0}{\bP} r +\frac{2\xbar\sigma_0}{\bP}(\xbar\Psi_1^0 + \xbar\phi_0^0\phi_1^0 - \tau_0) r^2 + O(r^3)] du dz\nn\\&&\hspace{1cm} + [4\frac{\tau_0}{P} r +\frac{2\sigma_0}{P}( \Psi_1^0 + \phi_0^0\xbar\phi_1^0 -\xbar\tau_0) r^2 + O(r^3)] dud\bz \nn\\&&\hspace{1cm}
 + [\frac{2\xbar\sigma_0}{\bP^2}r - \frac{4\xbar\Psi_0^0}{\bP^2}r^2 + O(r^3)] dz^2  + [\frac{2\sigma_0}{P^2}r  - \frac{4 \Psi_0^0}{P^2}r^2 + O(r^3)]d\bz^2 \nn\\&&\hspace{1cm}+ 2[-\frac{1}{P\bP} - \frac{2\sigma_0\xbar\sigma_0}{P\bP}r^2 + O(r^3)]dzd\bz,\label{nearmetric}\\
&&A_u=(\phi_1^0 + \xbar \phi_1^0) r + (\xbar\eth\phi_0^0 + \eth\xbar\phi_0^0) r^2 + O(r^3),\label{nearAu}\\
&&A_z=A_z^0 + \frac{\xbar\phi_0^0}{\bP} r + \half(\frac{\xbar\phi_0^1}{\bP}-\frac{\phi_0^0\sigma_0}{\bP})r^2 + O(r^3),\label{nearAz}
\eea
where $\gamma_0$ is a real constant indicating the surface gravity and $\sigma_0=\frac{1}{2\gamma_0}(\tau_0^2-\eth\tau_0)$.

\section{Asymptotic symmetries and conserved current near the isolated horizon}
The asymptotic behavior of our solution \eqref{nearmetric} is consistent with the choice of \cite{Donnay:2015abr}, which means the Bondi-like coordinate is compatible with such boundary condition. In the following, we will work out the asymptotic symmetry of Einstein-Maxwell theory near isolated horizon with the gauge and boundary condition given as
\bea
&&g_{rr}=g_{rz}=g_{r\bz}=A_r=0,\;\;g_{ur}=1,\label{gauge}\\
&&g_{uu}=4\gamma_0 r + O(r^2),\;\;g_{uz}=O(r),\;\;g_{u\bz}=O(r),\;\;\p_u g_{ab}=O(r^2),\label{boundary1}\\
&&A_u=O(r),\;\;A_z=O(1),\;\;A_{\bz}=O(1),\;\;\p_u A_\mu=O(r).\label{boundary2}
\eea

The gauge condition \eqref{gauge} will fix the asymptotic Killing vector and asymptotic $U(1)$ symmetry parameter $\zeta$ up to some integration constants
\begin{gather}
  \left\{\begin{array}{l}\label{presymmetry}
      \xi^u =f(u,z,\bz),\\
      \xi^r =R_r(u,z,\bz)-r\p_u f + \p_z f \int dr \;(g^{zz}g_{uz}+g^{z\bz}g_{u\bz}) + \p_{\bz} f \int dr \;(g^{\bz\bz}g_{u\bz}+g^{z\bz}g_{uz}),\\
      \xi^z =Y^z(u,z,\bz)-\p_z f \int dr \;g^{zz} - \p_{\bz} f \int dr \;g^{z\bz},\\
      \xi^{\bz} =Y^{\bz}(u,z,\bz)-\p_z f \int dr \;g^{z\bz} - \p_{\bz} f \int dr \;g^{\bz\bz},\\
      \zeta=\zeta_0(u,z,\bz) + \int dr\;(A_z \p_{\bz} f + A_{\bz} \p_z f).
\end{array}
\right.
\end{gather}
Then, the boundary condition \eqref{boundary1} and \eqref{boundary2} will fix the $u$ dependence of the integration constants. Hence, the asymptotic Killing vector and asymptotic $U(1)$ symmetry parameter $\zeta$ are further constrained to
\begin{gather}
  \left\{\begin{array}{l}\label{symmetry}
      \xi^u =T(z,\bz),\\
      \xi^r =\p_z T \int dr \;(g^{zz}g_{uz}+g^{z\bz}g_{u\bz}) + \p_{\bz} T \int dr \;(g^{\bz\bz}g_{u\bz}+g^{z\bz}g_{uz})=O(r^2),\\
      \xi^z =Y(z)-\p_z T \int dr \;g^{zz} - \p_{\bz} T \int dr \;g^{z\bz}=Y(z) + O(r),\\
      \xi^z =\bY(\bz)-\p_{\bz} T \int dr \;g^{\bz\bz} - \p_z T \int dr \;g^{z\bz}=\bY(\bz) + O(r),\\
      \zeta=\zeta(z,\bz) + \int dr\;(A_z \p_{\bz} T + A_{\bz} \p_z T)=\zeta(z,\bz) + O(r).
\end{array}
\right.
\end{gather}
Near horizon, the complete asymptotic symmetries of Einstein-Maxwell form a closed algebra as
\bea
&&[(\xi_1,\zeta_1),(\xi_2,\zeta_2)]_M=(\hat{\xi},\hat{\zeta}),\\
&&\hat{\xi}=[\xi_1 ,\xi_2],\;\;\hat{\zeta}=\xi_1(\zeta_2)-\xi_2(\zeta_1).
\eea
By turning off Maxwell fields, our asymptotic symmetry algebra will recover the one found in \cite{Donnay:2015abr} (see also \cite{laura}). Following the strategy of \cite{Barnich:2013axa}, one can compute the conserved current associated to the asymptotic symmetries and the current is derived by\footnote{We have set $8\pi G=c=1$ and the Lagrangian of Einstein-Maxwell theory is $\cL=\half \sqrt{g} (R+\half F^2)$.}
\bea\label{current}
\cJ^u_{\xi,\zeta}=-\frac{1}{P\bP}\big[2T\gamma_0 + Y[\frac{\xbar\tau_0}{\bP} + (\phi_1^0 + \xbar \phi_1^0)A_z^0] + \bY[\frac{\tau_0}{P} + (\phi_1^0 + \xbar \phi_1^0)A_{\bz}^0] + \zeta (\phi_1^0 + \xbar \phi_1^0)\big],
\eea
\bea
\cJ^z=0,
\eea
with a locally well-defined current algebra
\be\label{algebra}
\delta_{(\xi_2,\zeta_2)}\cJ^u_{(\xi_1,\zeta_1)}=\cJ^u_{(\hat{\xi},\hat{\zeta})} + \p_a L^{[u a]}_{(\xi_1,\zeta_2),(\xi_2,\zeta_2)},
\ee
while
\be
L^{[u z]}_{(\xi_1,\zeta_2),(\xi_2,\zeta_2)}=Y_2 \cJ^u_{(\xi_1,\zeta_1)}.
\ee
The asymptotic current can be adapted into a more consistent way
\be
\cJ^u_{\xi,\zeta}=\frac{1}{P\bP} \;\;\big[\xi \cdot \mathbf{\omega} + \xi \cdot\mathbf{\cA}(\phi_1^0 + \xbar \phi_1^0)+ \zeta (\phi_1^0 + \xbar \phi_1^0)\big],
\ee
where the extrinsic curvature one-form on the horizon is defined by
\be
\omega_a\he l_b\nabla_a n^b\he -2\gamma_0 l_a + \tau_0 \xbar m_a + \xbar\tau_0 m_a\nn
\ee
and $\xi\he Tn+\cY m + \xbar\cY \xbar m$ is the asymptotic Killing vector while
\be
\cY=\frac{\bY(\bz)}{P},\;\;\xbar\cY=\frac{Y(z)}{\bP},\;\;\mathbf{\cA}=P A_{\bz}^0\xbar m + \bP A_{z}^0 m.\nn
\ee

Amazingly, the current associated with supertranslation gets no contribution from electromagnetic fields. This reveals the fact that the electromagnetic fields become soft on the isolated horizon and do not contribute to black hole energy which is consistent with the statement in \cite{Hawking:2016msc}. The superrotation parts indeed have been modified from the polarization of the soft photons.

\section{Soft electric hairs on isolated horizon}

In this section we will deal with a special catalogue of solution with more concrete physical interpretation. The isolated horizon is spherically symmetric and has unit radius, \textit{i.e.} $P=\bP=\frac{1}{\sqrt{2}}(1+z\bz)$ and $\tau_0=0$. Electromagnetic fields are not vanishing near the horizon. This case can be considered as the later state of electromagnetic fields around a Schwarzschild black hole with mass $M=\half$. Back-scattered radiation can be neglected but wave can still radiate in the region far away from the horizon, eventually will be scattered at null infinity. This specification will lead to a globally well-defined horizon charge from the conserved current respecting to asymptotic $U(1)$ symmetry and supertranslation. The charge associated to asymptotic $U(1)$ symmetry will be deduced to
\bea\label{softcharge}
\cQ^{\cH}_\zeta=\int_{S^2} d\Omega^2 \zeta(\phi_1^0 + \xbar \phi_1^0),
\eea
where $d\Omega^2$ is the unit spherical surface element. By expanding the asymptotic $U(1)$ symmetry parameter $\zeta$ in spherical harmonics
\bea
\zeta=\sum\limits_{h=0}^\infty \sum\limits_{g=-h}^h \zeta_{h,g}Y_{h,g}(z,\bz),\nn
\eea
one can further define the modes of the horizon charge as
\bea\label{softmodes}
\cQ^{\cH}_{\zeta_{h,g}}=\int_{S^2} d\Omega^2 (\phi_1^0 + \xbar \phi_1^0)Y_{h,g}.
\eea

Interestingly, the modes we have defined in \eqref{softmodes} are equivalent to the electric multipoles introduced in \cite{Ashtekar:2004gp}. The supertranslation charges will vanish except the zero mode which becomes a combination proportional to the surface gravity multiplying the horizon area \cite{Donnay:2015abr,Eling:2016xlx}. This also has a counterpart from the mass multipoles in \cite{Ashtekar:2004gp}, where the mass monopole $M_0$ is the only non-zero mass multipoles of a spherically symmetric isolated horizon. Thus, one immediately application of the soft electric hairs is to capture the electric multipoles information on the horizon\footnote{In \cite{Ashtekar:2004nd}, multipoles are served as required parameters in the construction of the ensemble to calculate the black hole entropy.}. This indeed confirms that the later stage of a black hole collapse carries infinite numbers of soft electric hairs. As we have shown in the previous section that the electromagnetic fields become soft on the isolated horizon, the modes \eqref{softmodes} can be understood as soft photons located on the horizon during the dynamical process before the isolated horizon formed. As shown in Section \ref{soluchapter}, the Maxwell fields are fully characterized by the electromagnetic fields $A^0_z$ and the real part of $\phi^0_1$. To track the whole information of the electromagnetic fields, one still needs to consult to the local information from the superrotation current.

The isolated horizon is admitted by a black hole who itself is in equilibrium but whose exterior contains radiation (\textit{i.e.} the whole spacetime is not yet stationary). The huge amount of classical charges we have introduced on the isolated horizon have nothing to violate the no hair theorem \cite{Chrusciel:2012jk} because it is only valid when the whole spacetime is stationary. One may wonder whether the charges will act on a quantum state trivially or not, which would be equivalent to asking the asymptotic $U(1)$ symmetry is spontaneously broken or not. As far as we can understand, it's not necessary to have the asymptotic $U(1)$ symmetry spontaneously broken. The reason is that the system is already in equilibrium around the isolated horizon. There will be no longer photon or soft photon reaching the horizon. Thus, the system has no interaction and becomes a free theory. Hence, the quantum states are supposed to be the engine states of the quantized operators $\cQ^{\cH}_{h,g}$. This can be also seen from the fact that the horizon charges, we have defined in \eqref{softcharge}, do not include a soft part compared to the one defined in \cite{He:2014cra}. However, this symmetry will be broken by the appearance of radiation crossing the horizon. It will be highly interesting to investigate the charges on a dynamical horizon \cite{Ashtekar:2003hk}. One could expect that it will be very similar to the case of null infinity where the asymptotic charges have both the hard part and the soft part. Accordingly these charges will act on quantum states non-trivially.

\section{Supertranslation and foliation of an isolated horizon}
We would like to have a deeper investigation on the action of the supertranslation on the horizon. A supertranslation will preserve the induced horizon metric, which is degenerate, and the null normal vector $n$. But the extrinsic curvature one-form will be transformed like the gradient of a scalar
\bea\label{supertransf}
\omega_a'\rightarrow \omega_a - 2\gamma_0 d f.
\eea
Such a transformation is related to the ambiguity of the foliation of an isolated horizon which was discussed in\cite{Ashtekar:2001jb,Ashtekar:2004gp}.

Let's consider a non-extremal($\gamma_0\neq0$) isolated horizon ($\cH,n$). We will ignore the Maxwell fields and focus only on the horizon geometry determined by the induced metric $q_{ab}$ and the induced derivative operator $D$ in this section. Then a fixed cross-section $S$ of $\cH$ can be treated as a leaf of a foliation $u=$constant such that $n^aD_a u=1$ and the normal $l_a$ of this foliation can be set as $l_a=D_a u$ with $l_an^a=1$. A projection operator $\tilde{q}^b_a$ on the leaves of the foliation is defined by $\tilde{q}^b_a=\delta^b_a - l_an^b$. Since $q_{ab}$ is degenerate, $D$ can not be fully determined by $q_{ab}$. But on the cross-section $S$, one has a unique (torsion-free) derivative operator $\hat{D}$ compatible with $\hat{q}_{ab}$ the projection of $q_{ab}$ on $S$. To determine the derivative operator $D$ on the horizon, one only needs to specify its action on $l_a$\footnote{The action of $D$ on $n^a$ is $D_an^b\he\omega_an^b$.}. Let's define $S_{ab}:=D_al_b$ who satisfies $S_{ab}n^b\he-\omega^a$ on the horizon. Then the horizon geometry is completely specified by the triplet ($q_{ab},\omega_a,S_{ab}$).

Suppose the triplet ($q_{ab},\omega_a,S_{ab}$) is given on the horizon with one foliation $u=$constant, hence on cross-section $S$, the free data ($\hat{q}_{ab},\hat{\omega}_a,\hat{S}_{ab}$) can be derived by the projection operator from the horizon $\cH$ as
\bea
&&\hat{q}_{ab}=q_{ab},\\
&&\hat{\omega}_a=\omega_a + 2\gamma_0l_a,\\
&&\hat{S}_{ab}=S_{ab}+l_a\omega_b+\omega_al_b+2\gamma_0l_al_b.
\eea
Let's now consider another cross section $S'$ which does not belong to the same foliation. One can choose $u'=$constant as the corresponding foliation. Let $f=u'-u$ and $\cL_nf=0$. The two sets of free data are related by
\bea
&&\hat{q}_{ab}=\hat{q'}_{ab},\\
&&\hat{\omega}_a=\hat{\omega'}_a-2\gamma_0df,\\
&&\hat{S}_{ab}=\hat{S'}_{ab}+D_a D_b f.
\eea
This is the ambiguity of choosing foliation of an isolated horizon. This ambiguity can be also understood in an inverse way that the same set of free data can create different foliation. The difference of $\omega_a$ matches precisely a supertranslation transformation on the horizon given in \eqref{supertransf} in the beginning of this section.

The study of null infinity reveals an fascinating relation between the supertranslation and the memory effect. It seems quite promising that there should exist the analogical memory effect on the horizon. The memory effect is non-zero change of asymptotic shear, which is caused by some dynamical processes. Such result has been realized by a supertranslation at null infinity recently in \cite{Strominger:2014pwa}. Similar things also happen on a horizon. As discussed previously, the foliation of an isolated horizon has some ambiguities and different foliations are related by supertranslation on the horizon. A foliation of an isolated horizon can be connected to another one by a dynamical process before it formed. Since the foliation of dynamical horizon is unique\cite{Ashtekar:2005ez}, the final foliation of isolated horizon is fixed by continuity condition, \textit{i.e.} different foliation corresponds to different dynamical process of black hole. This is quite similar to what happens at null infinity. It would be definitely worthwhile to investigate such an effect elsewhere.

\section{Discussion}
We hope to have convinced the readers interested in the recent proposal in \cite{Hawking:2016msc} that the near horizon physics is much richer then what we have understood. We have shown precisely that there are actually a large amount of soft electric hairs on the isolated horizon. Now we would like to comment on the supertranslation hairs and superrotation hairs. We will focus on pure gravity case. But the way of taking consider of the contribution from electromagnetic fields is obvious. The main difference between null infinity and isolated horizon in asymptotic analysis is the fact that null infinity is a conformal object. Hence, null infinity can be always set to $S^2\times R$. This choice will not break superrotation since the conformal transformation can be compensated by rescaling in $r$. But isolated horizon has to be $\mathbb{S}\times R$ to include superrotation as near horizon symmetries where $\mathbb{S}$ is conformally $S^2$. The difficulty of defining supertranslation or superrotation modes are from the fact that a complete basis is not known for expansion on a generic 2-manifold. In practice, we just integrate the current \eqref{current} on a unit sphere\footnote{Integration on Riemann sphere was used in \cite{Donnay:2015abr}.}
\be
\cQ_\xi=\int \td \Omega^2\;2T\gamma_0\Theta + Y\Xi \Theta + \bY\xbar\Xi \Theta,
\ee
where $\xbar\Xi=\frac{\tau_0}{P},\Xi=\frac{\xbar\tau_0}{\bP}$ and $\Theta=-\frac{(1+z\bz)^2}{2P\bP}$. Essentially, the supertranslation and superrotation hairs are indicating information of the geometry (\textit{i.e.} the intrinsic and extrinsic curvature of $\mathbb{S}$) and topology (\textit{i.e.} punctures on $\mathbb{S}$) of the isolated horizon. The supertranslation and superrotation hairs would be related to the geometric multipoles defined in \cite{Ashtekar:2004gp}. Because the near horizon symmetry algebra \eqref{algebra} is not Abelian, soft hairs were shown to be implanted by finite transformation in \cite{Hawking:2016sgy}.

Another very meaningful open question is that if those soft hairs can be observed from the region not near horizon. Alternatively we are asking if the near horizon symmetry can be extended to the region not very close to the horizon. Recent investigations \cite{Compere:2015mza,Compere:2015bca,Compere:2015knw,Sheikh-Jabbari:2016lzm} on three dimensional pure Einstein gravity has enhanced
the asymptotic symmetry to symplectic symmetry which allows one to define conserved charges anywhere in the bulk (see also \cite{Seraj:2016jxi} for pure Maxwell case). The crucial step to promote asymptotic symmetry to symplectic symmetry was observed in \cite{Compere:2014cna} that the asymptotic conserved charges are $r$-independent. The solution space \eqref{nearmetric}-\eqref{nearAz} allows us to check the electric charge easily and the charge is still $r$-dependent. However we still have hope in extending the near horizon symmetry to the region away from the horizon (the emergence of symplectic symmetry would be expected). As shown in \cite{Conde:2016csj,Conde:2016rom}, the sub-leading pieces of the charge can have their own physical effects when one acts the charge on a quantum state. Naively, the subleading contributions can be understood as the effect of local degree of freedom in the bulk in the case that it is not very close to the boundary. In particular, the vanishing of the subleading pieces of the charge in \cite{Compere:2014cna} is a consequence of the absence of local degree of freedom in three dimensional pure Einstein gravity. It would be very interesting to investigate the physical effect of the subleading pieces of the charge in the near horizon case elsewhere.

\section*{Acknowledgments}

This work is partially supported by the National Natural Science
Foundation of China (Grant Nos. 11575286, 11475179, and 11675015), by the Fund for Scientific
Research-FNRS (Belgium), by IISN-Belgium, and by ``Communaut\'e fran\c caise de Belgique - Actions de Recherche Concert\'ees''.
P.~Mao is supported by a PhD fellowship from the China Scholarship Council. He thanks G. Barnich, L. Donnay, G. Giribet, H. Gonz\'{a}lez, R.Li, J. Long, and Y. Zhang for useful discussions. H.~Zhang is supported in part by the Belgian Federal Science Policy
Office through the Interuniversity Attraction Pole P7/37, by FWO-Vlaanderen through the project G020714N, by the Vrije Universiteit Brussel through the Strategic Research Program ``High-Energy Physics''. He is also an individual FWO Fellow supported by 12G3515N.

\appendix

\section{Solution space}\label{solution}
In a Bondi-like coordinate ($u,r,z,\bz$), the standard Newman-Penrose prescription can always have the following ansatz:
\bea
&&l^\mu=[0,1,0,0],\;\;n^\mu=[1,U,X^A],\;\;m^\mu=[0,\omega,L^A].\nn\\
&&l_\mu=[1,0,0,0],\;\;n_\mu=[-U-X^A(\bomega L_A+\omega \bL_A),1,\omega\bL_A+\bomega L_A],\nn\\
&&m_\mu=[-X^AL_A,0,L_A].\nn\\
&&\kappa=\pi=\epsilon=0,\;\;\rho=\bar{\rho},\;\;\tau=\bar{\alpha}+\beta.\nn
\eea
where $L^AL_A=0,L^A\bL_A=-1$. Under this gauge choice, the Newman-Penrose equations can be solved out recursively. We would refer directly the number of equations in \cite{Chandrasekhar}. Firstly, the solutions on the horizon can be worked out through hypersurface equations (305), (306), (310.j)-(310.r), Bianchi identities (321.f)-(321.h) and Maxwell equations (333). Then, apart from $\Psi_0,\phi_0$ who have to be given at any order of $r$ as initial data, the asymptotic $r$ dependence of all the rest variables can be calculated by the radial equations (303), (304), (309), (310.a)-(310.r), Bianchi identities (321.a)-(321.d) and Maxwell equations (330)-(331). Finally, (324) will determine Maxwell potential $A_\mu$ under the gauge condition we have chosen.

The full solutions are listed up to order $O(r^3)$ as following:
\bea
&&\gamma=\gamma_0 + \gamma_1 r + \gamma_2 r^2 + O(r^3),\;\;\mbox{$\gamma_0$ is a real constant.}\\
&&\;\;\;\;\;\;\;\;\gamma_1=\alpha_0\tau_0 + \beta_0\xbar\tau_0 + \Psi_2^0 + \phi_1^0\xbar\phi_1^0,\nn\\
&&\;\;\;\;\;\;\;\;\gamma_2=\half(\tau_0\alpha_1+\tau_1\alpha_0+\xbar\tau_0\beta_1+\xbar\tau_1\beta_0+\Psi_2^1+\phi_1^0\xbar\phi_1^0+\phi_1^1\xbar\phi_1^0) ,\nn\\
&&\tau=\tau_0 + \tau_1 r + \tau_2 r^2 + O(r^3),\;\;\tau_1=\Psi_1^0 + \phi_0^0\xbar\phi_1^0,\\
&&\;\;\;\;\;\;\;\;\tau_2=\half (\tau_0\rho_1+\tau_1\rho_0+\xbar\tau_0\sigma_1+\xbar\tau_1\sigma_0\Psi_1^1+\phi_0^1\xbar\phi_0^0\phi_0^0\xbar\phi_0^1),\nn\\
&&\rho=\rho_0 + \rho_1 r + \rho_2 r^2 + O(r^3),\\
&&\;\;\;\;\;\;\;\;\rho_0=\frac{1}{4\gamma_0}[\Psi_2^0 + \xbar \Psi_2^0 - \eth \xbar \tau_0 - \xbar\eth\tau_0 + 2 \tau_0\xbar\tau_0],\;\;\rho_1=\phi_0^0\xbar\phi_0^0,\nn\\
&&\;\;\;\;\;\;\;\;\rho_2=\half(\sigma_0\xbar\sigma_1+\sigma_1\xbar\sigma_0+2\rho_0\rho_1+\phi_0^1\xbar\phi_0^0+\phi_0^0\xbar\phi_0^1) ,\nn\\
&&\sigma=\sigma_0 + \sigma_1 r + \sigma_2 r^2 + O(r^3),\;\;\sigma_0=\frac{1}{2\gamma_0}[\tau_0^2-\eth\tau_0],\;\;\sigma_1=\Psi_0^0,\\
&&\;\;\;\;\;\;\;\;\sigma_2=\half(2\sigma_0\rho_1+2\sigma_1\rho_0+\Psi_0^1),\nn\\
&&\alpha=\alpha_0 + \alpha_1 r + \alpha_2 r^2 + O(r^3),\;\;\alpha_0=\half(\xbar\tau_0 + \xbar\delta\ln P),\;\;\alpha_1=\phi_1^0\xbar\phi_0^0,\\
&&\;\;\;\;\;\;\;\;\alpha_2=\half(\rho_0\alpha_1+\rho_1\alpha_0+\xbar\sigma_0\beta_1+\xbar\sigma_1\beta_0+\phi_1^1\xbar\phi_0^0+\phi_1^0\xbar\phi_0^1),\nn\\
&&\beta=\beta_0 + \beta_1 r + \beta_2 r^2 + O(r^3),\;\;\beta_0=\half(\tau_0 - \delta\ln \bP),\;\;\beta_1=\Psi_1^0,\\
&&\;\;\;\;\;\;\;\;\beta_2=\half(\alpha_0\sigma_1+\alpha_1\sigma_0+\Psi_1^1),\nn\\
&&\mu=\Psi_2^0 r + \half\Psi_2^1 r^2 + O(r^3),\\
&&\lambda=\half(\mu_1\xbar\sigma_0+\phi_2^1\xbar\phi_0^0)r^2 + O(r^3),\\
&&\nu=\half(\mu_1\xbar\tau_0+\Psi_3^1+\phi_2^1\xbar\phi_1^0)r^2 + O(r^3),\\
&&\phi_0=\phi_0^0 + \phi_0^1 r + O(r^2),\\
&&\phi_1=\phi_1^0 + \phi_1^1 r + \phi_1^2 r^2 +  O(r^3),\;\;\phi_1^1=\xbar\eth\phi_0^0 - \tau_0 \phi_0^0,\\
&&\;\;\;\;\;\;\;\;\phi_1^2=\half(\xbar\eth\phi_0^1-\xbar\tau_0\phi_0^1-2\alpha_1\phi_0^0+2\rho_1\phi_1^0),\nn\\
&&\phi_2= \xbar\eth\phi_2^0 r + \half \xbar\eth\phi_1^1 r^2 +  O(r^3),\\
&&\Psi_0=\Psi_0^0 + \Psi_0^1 + O(r^2),\\
&&\Psi_1=\Psi_1^0 + \Psi_1^1 r  + \Psi_1^2 r^2 + O(r^3),\\
&&\;\;\;\;\;\;\;\;\Psi_1^0=\xbar\eth\sigma_0 - \eth \rho_0 -\sigma_0\xbar\tau_0+\rho_0\tau_0 + \phi_0^0\xbar\phi_1^0,\nn\\
&&\;\;\;\;\;\;\;\;\Psi_1^1=\xbar\eth\sigma_1 - \eth \rho_1 -\sigma_1\xbar\tau_0+\rho_1\tau_0  -\sigma_0\xbar\tau_1+\rho_0\tau_1 + \phi_0^1\xbar\phi_1^0 + \phi_0^0\xbar\phi_1^1,\nn\\
&&\;\;\;\;\;\;\;\;\Psi_1^2=\xbar\eth\sigma_2 - \eth \rho_2 +\rho_0\tau_2+\rho_2\tau_0+\rho_1\tau_1 -2(\alpha_2-\xbar\beta_2)\sigma_0 -2 (\alpha_1-\xbar\beta_1)\sigma_1,\nn\\
&&\;\;\;\;\;\;\;\;\;\;\;\;\;\;=-\xbar\tau_2\sigma_0-\xbar\tau_1\sigma_1-\xbar\tau_0\sigma_2+\phi_0^0\xbar\phi_1^2+\phi_0^1\xbar\phi_1^1+\phi_0^2\xbar\phi_1^0,\nn\\
&&\Psi_2=\Psi_2^0 + \Psi_2^1 r + \Psi_2^2 r^2 + O(r^3),\\
&&\;\;\;\;\;\;\;\;\Psi_2^0=\phi_1^0\xbar\phi_1^0 - \frac{R}{4} +\half(\xbar\eth\tau_0 - \eth\xbar\tau_0),\nn\\
&&\;\;\;\;\;\;\;\;\Psi_2^1=\phi_1^1\xbar\phi_1^0 + \phi_1^0\xbar\phi_1^1 + \xbar\eth\beta_1 - \eth \alpha_1 + \alpha_0 \xbar\alpha_1 + \beta_0 \xbar \beta_1 - \alpha_0 \beta_1 -\alpha_1 \beta_0,\nn\\
&&\;\;\;\;\;\;\;\;\Psi_2^2=\xbar\eth\beta_2 - \eth \alpha_2 +\rho_1\mu_1+\rho_0\mu_2 - \sigma_0\lambda_2 +\alpha_0\xbar\alpha_2+\alpha_1\xbar\alpha_1+\beta_0\xbar\beta_2+\beta_1\xbar\beta_1,\nn\\
&&\;\;\;\;\;\;\;\;\;\;\;\;\;\;=-\alpha_0\xbar\beta_2-\alpha_2\beta_0-2\alpha_1\beta_1+\phi_1^0\xbar\phi_1^2+\phi_1^1\xbar\phi_1^1+\phi_1^2\xbar\phi_1^0,\nn\\
&&\Psi_3=\Psi_3^1 r + \Psi_3^2 r^2 + O(r^3),\;\;\Psi_3^1=\xbar\eth\mu_1 + \mu_1 \xbar\tau_0 + \phi_2^1\xbar\phi_1^0,\\
&&\;\;\;\;\;\;\;\;\Psi_3^2=\xbar\eth\mu_2-\eth\lambda_2+\mu_1\xbar\tau_1+\mu_2\xbar\tau_0-\tau_0\lambda_2+\phi_2^1\xbar\phi_1^1+\phi_2^2\xbar\phi_1^0,\nn\\
&&\Psi_4=\Psi_4^2 r^2 + O(r^3),\\
&&\;\;\;\;\;\;\;\;\Psi_4^2=\xbar\eth\nu_2-\half\phi_2^1(\xbar\eth\xbar\phi_1^0-2\xbar\tau_0\xbar\phi_1^0+2\gamma_0\xbar\phi_0^0)-2\gamma_0\lambda_2+\xbar\tau_0\nu_2,\nn\\
&&X^z=\tau_0 \bP r + \half (\tau_1\bP+\xbar\tau_0\sigma_0\bP)r^2 + O(r^3),\\
&&X^{\bz}=\xbar\tau_0 P r + \half (\xbar\tau_1 P+ \tau_0\xbar\sigma_0 P)r^2 + O(r^3),\\
&&\omega=-\tau_0 r + \half(\sigma_0\xbar\omega_0-\tau_1)r^2 + O(r^3),\\
&&U=-2 \gamma_0 r + \half(\tau_0\xbar\omega_1+\xbar\tau_0\omega_1-\gamma_1-\xbar\gamma_1) r^2 + O(r^3),\\
&&L^z=\sigma_0 \bP r + \half \sigma_1\bP r^2 +  O(r^3),\;\;L^{\bz}=P + \half \sigma_0\xbar\sigma_0 P r^2 + O(r^3),\\
&&\xbar L^z=\bP + \half \xbar\sigma_0\sigma_0 \bP r^2 + O(r^3),\;\;\xbar L^{\bz}=\xbar\sigma_0 P r + \half \xbar\sigma_1 P r^2 +  O(r^3),\\
&&L_z=-\dfrac{1}{\bP} -\frac{\sigma_0\xbar\sigma_0}{2\bP} r^2+ O(r^3),\;\;L_{\bz}=\frac{\sigma_0}{P} r + \frac{\sigma_1}{2P}r^2 + O(r^3),\\
&&\xbar L_z=\frac{\xbar\sigma_0}{\bP} r + \frac{\xbar\sigma_1}{2\bP}r^2 + O(r^3),\;\;\xbar L_{\bz}=-\frac{1}{P} -\frac{\sigma_0\xbar\sigma_0}{2P} r^2+ O(r^3),\\
&&A_u=(\phi_1^0 + \xbar \phi_1^0) r + (\phi_1^1 + \xbar \phi_1^1+\tau_0\phi_0^0+\xbar\tau_0\xbar\phi_0^0) r^2 + O(r^3),\\
&&A_z=A_z^0 +\frac{\xbar\phi_0^0}{\bP} r + \half(\frac{\xbar\phi_0^1}{\bP}-\frac{\phi_0^0\sigma_0}{\bP}) r^2 + O(r^3),\;\;\p_{\bz}A_z^0-\p_{z}A_{\bz}^0=\frac{\xbar\phi_1^0 - \phi_1^0}{P\bP},\nn
\eea
where $\tau_0,P,\phi_1^0,A_z^0$ are arbitrary functions depending on $(z,\bz)$ only, and $R=2(\delta\xbar\delta\ln P + \xbar\delta\delta\ln\bP-2\xbar\delta\ln P\delta\ln\bP)=2 \bP P \p_z\p_{\bz} \ln\bP P$ is the scalar curvature of $\mathbb{S}$. $\tau_0$ is the extrinsic curvature of the $\mathbb{S}$ on the horizon. The $\eth$ operator is defined by $\eth\eta=\delta\eta + s \delta \ln \bP \eta$ and $\xbar\eth\eta=\xbar\delta\eta - s \xbar\delta \ln P \eta$ where $s$ is the spin weight of the field $\eta$ and the spin weight of relevant fields are listed in Table \ref{t1}. The time evolution of $\Psi_0$ and $\phi_0$ are fully controlled by (321.e) and (332).
\begin{table}
\caption{Spin weights}\label{t1}
\begin{center}\begin{tabular}{c|c|c|c|c|c|c|c|c|c|c|c|c|c|c|c|c|c} & $\eth$ & $\tau$ & $\alpha$ & $\beta$ & $\rho$ & $\gamma$ & $\nu$ & $\mu$ & $\sigma$ & $\lambda$  & $\Psi^0_4$ &  $\Psi^0_3$ & $\Psi^0_2$ & $\Psi^0_1$ & $\Psi_0^0$ & $\cY$   \\
\hline
s & $1$& $1$& $-1$& $1$& $0$& $0$& $-1$& $0$& $2$& $-2$  &
  $-2 $&  $-1$ & $0$ & $1$ & $2$ & $-1$  \\
\end{tabular}\end{center}\end{table}

\bibliography{refs1,new}

\providecommand{\href}[2]{#2}\begingroup\raggedright\begin{thebibliography}{10}

\bibitem{Bondi:1962px}
H.~Bondi, M.~G.~J. van~der Burg, and A.~W.~K. Metzner, ``{Gravitational waves
  in general relativity. 7. Waves from axisymmetric isolated systems},''
\href{http://dx.doi.org/10.1098/rspa.1962.0161}{{\em Proc. Roy. Soc. Lond.}
  {\bfseries A269} (1962) 21--52}.
%%CITATION = PRSLA,A269,21;%%.

\bibitem{Sachs:1962wk}
R.~K. Sachs, ``{Gravitational waves in general relativity. 8. Waves in
  asymptotically flat space-times},''
\href{http://dx.doi.org/10.1098/rspa.1962.0206}{{\em Proc. Roy. Soc. Lond.}
  {\bfseries A270} (1962) 103--126}.
%%CITATION = PRSLA,A270,103;%%.

\bibitem{Sachs:1962zza}
R.~Sachs, ``{Asymptotic symmetries in gravitational theory},''
\href{http://dx.doi.org/10.1103/PhysRev.128.2851}{{\em Phys. Rev.} {\bfseries
  128} (1962) 2851--2864}.
%%CITATION = PHRVA,128,2851;%%.

\bibitem{Newman:1966ub}
E.~T. Newman and R.~Penrose, ``{Note on the Bondi-Metzner-Sachs group},''
\href{http://dx.doi.org/10.1063/1.1931221}{{\em J. Math. Phys.} {\bfseries 7}
  (1966) 863--870}.
%%CITATION = JMAPA,7,863;%%.

\bibitem{Mccarthy:1972ry}
P.~J.~M. Mccarthy, ``{Asymptotically flat space-times and elementary
  particles},''
\href{http://dx.doi.org/10.1103/PhysRevLett.29.817}{{\em Phys. Rev. Lett.}
  {\bfseries 29} (1972) 817--819}.
%%CITATION = PRLTA,29,817;%%.

\bibitem{Crampin:1974aw}
M.~Crampin, ``{Physical significance of the topology of the bondi-metzner-sachs
  group},''
\href{http://dx.doi.org/10.1103/PhysRevLett.33.547}{{\em Phys. Rev. Lett.}
  {\bfseries 33} (1974) 547--550}.
%%CITATION = PRLTA,33,547;%%.

\bibitem{Girardello:1974sq}
L.~Girardello and G.~Parravicini, ``{Continuous spins in the
  bondi-metzner-sachs group of asymptotic symmetry in general relativity},''
\href{http://dx.doi.org/10.1103/PhysRevLett.32.565}{{\em Phys. Rev. Lett.}
  {\bfseries 32} (1974) 565--568}.
%%CITATION = PRLTA,32,565;%%.

\bibitem{Piard:1977vh}
A.~Piard, ``{Representations of the Bondi-Metzner-Sachs Group with the Hilbert
  Topology},''
\href{http://dx.doi.org/10.1016/0034-4877(77)90068-4}{{\em Rept. Math. Phys.}
  {\bfseries 11} (1977) 279--283}.
%%CITATION = RMHPB,11,279;%%.

\bibitem{Barnich:2009se}
G.~Barnich and C.~Troessaert, ``{Symmetries of asymptotically flat 4
  dimensional spacetimes at null infinity revisited},''
  \href{http://dx.doi.org/10.1103/PhysRevLett.105.111103}{{\em Phys. Rev.
  Lett.} {\bfseries 105} (2010) 111103},
\href{http://arxiv.org/abs/0909.2617}{{\ttfamily arXiv:0909.2617 [gr-qc]}}.
%%CITATION = ARXIV:0909.2617;%%.

\bibitem{Barnich:2010eb}
G.~Barnich and C.~Troessaert, ``{Aspects of the BMS/CFT correspondence},''
  \href{http://dx.doi.org/10.1007/JHEP05(2010)062}{{\em JHEP} {\bfseries 05}
  (2010) 062},
\href{http://arxiv.org/abs/1001.1541}{{\ttfamily arXiv:1001.1541 [hep-th]}}.
%%CITATION = ARXIV:1001.1541;%%.

\bibitem{Barnich:2011mi}
G.~Barnich and C.~Troessaert, ``{BMS charge algebra},''
  \href{http://dx.doi.org/10.1007/JHEP12(2011)105}{{\em JHEP} {\bfseries 12}
  (2011) 105},
\href{http://arxiv.org/abs/1106.0213}{{\ttfamily arXiv:1106.0213 [hep-th]}}.
%%CITATION = ARXIV:1106.0213;%%.

\bibitem{Barnich:2016lyg}
G.~Barnich and C.~Troessaert, ``{Finite BMS transformations},''
  \href{http://dx.doi.org/10.1007/JHEP03(2016)167}{{\em JHEP} {\bfseries 03}
  (2016) 167},
\href{http://arxiv.org/abs/1601.04090}{{\ttfamily arXiv:1601.04090 [gr-qc]}}.
%%CITATION = ARXIV:1601.04090;%%.

\bibitem{Compere:2016jwb}
G.~Comp\`{e}re and J.~Long, ``{Vacua of the gravitational field},''
  \href{http://dx.doi.org/10.1007/JHEP07(2016)137}{{\em JHEP} {\bfseries 07}
  (2016) 137},
\href{http://arxiv.org/abs/1601.04958}{{\ttfamily arXiv:1601.04958 [hep-th]}}.
%%CITATION = ARXIV:1601.04958;%%.

\bibitem{Gervais:1976ec}
J.-L. Gervais, B.~Sakita, and S.~Wadia, ``{The Surface Term in Gauge
  Theories},''
\href{http://dx.doi.org/10.1016/0370-2693(76)90467-6}{{\em Phys. Lett.}
  {\bfseries B63} (1976) 55}.
%%CITATION = PHLTA,B63,55;%%.

\bibitem{Regge:1974zd}
T.~Regge and C.~Teitelboim, ``{Role of Surface Integrals in the Hamiltonian
  Formulation of General Relativity},''
\href{http://dx.doi.org/10.1016/0003-4916(74)90404-7}{{\em Annals Phys.}
  {\bfseries 88} (1974) 286}.
%%CITATION = APNYA,88,286;%%.

\bibitem{Benguria:1976in}
R.~Benguria, P.~Cordero, and C.~Teitelboim, ``{Aspects of the Hamiltonian
  Dynamics of Interacting Gravitational Gauge and Higgs Fields with
  Applications to Spherical Symmetry},''
\href{http://dx.doi.org/10.1016/0550-3213(77)90426-6}{{\em Nucl. Phys.}
  {\bfseries B122} (1977) 61}.
%%CITATION = NUPHA,B122,61;%%.

\bibitem{Ashtekar:1978zz}
A.~Ashtekar and R.~O. Hansen, ``{A unified treatment of null and spatial
  infinity in general relativity. I - Universal structure, asymptotic
  symmetries, and conserved quantities at spatial infinity},''
\href{http://dx.doi.org/10.1063/1.523863}{{\em J. Math. Phys.} {\bfseries 19}
  (1978) 1542--1566}.
%%CITATION = JMAPA,19,1542;%%.

\bibitem{Geroch:1977jn}
R.~Geroch, ``{Asymptotic structure of space-time},'' in {\em {Asymptotic
  structure of space-time}}, {P. Esposito and L. Witten}, ed., ch.~1,
  pp.~1--105.
\newblock Plenum, New York, 1977.

\bibitem{Brown:1986nw}
J.~D. Brown and M.~Henneaux, ``{Central Charges in the Canonical Realization of
  Asymptotic Symmetries: An Example from Three-Dimensional Gravity},''
\href{http://dx.doi.org/10.1007/BF01211590}{{\em Commun. Math. Phys.}
  {\bfseries 104} (1986) 207--226}.
%%CITATION = CMPHA,104,207;%%.

\bibitem{Strominger:1997eq}
A.~Strominger, ``{Black hole entropy from near horizon microstates},''
  \href{http://dx.doi.org/10.1088/1126-6708/1998/02/009}{{\em JHEP} {\bfseries
  02} (1998) 009},
\href{http://arxiv.org/abs/hep-th/9712251}{{\ttfamily arXiv:hep-th/9712251
  [hep-th]}}.
%%CITATION = HEP-TH/9712251;%%.

\bibitem{Strominger:2013jfa}
A.~Strominger, ``{On BMS Invariance of Gravitational Scattering},''
  \href{http://dx.doi.org/10.1007/JHEP07(2014)152}{{\em JHEP} {\bfseries 07}
  (2014) 152},
\href{http://arxiv.org/abs/1312.2229}{{\ttfamily arXiv:1312.2229 [hep-th]}}.
%%CITATION = ARXIV:1312.2229;%%.

\bibitem{He:2014laa}
T.~He, V.~Lysov, P.~Mitra, and A.~Strominger, ``{BMS supertranslations and
  Weinberg's soft graviton theorem},''
  \href{http://dx.doi.org/10.1007/JHEP05(2015)151}{{\em JHEP} {\bfseries 05}
  (2015) 151},
\href{http://arxiv.org/abs/1401.7026}{{\ttfamily arXiv:1401.7026 [hep-th]}}.
%%CITATION = ARXIV:1401.7026;%%.

\bibitem{Weinberg:1965nx}
S.~Weinberg, ``{Infrared photons and gravitons},''
\href{http://dx.doi.org/10.1103/PhysRev.140.B516}{{\em Phys. Rev.} {\bfseries
  140} (1965) B516--B524}.
%%CITATION = PHRVA,140,B516;%%.

\bibitem{He:2014cra}
T.~He, P.~Mitra, A.~P. Porfyriadis, and A.~Strominger, ``{New Symmetries of
  Massless QED},'' \href{http://dx.doi.org/10.1007/JHEP10(2014)112}{{\em JHEP}
  {\bfseries 10} (2014) 112},
\href{http://arxiv.org/abs/1407.3789}{{\ttfamily arXiv:1407.3789 [hep-th]}}.
%%CITATION = ARXIV:1407.3789;%%.

\bibitem{Strominger:2013lka}
A.~Strominger, ``{Asymptotic Symmetries of Yang-Mills Theory},''
  \href{http://dx.doi.org/10.1007/JHEP07(2014)151}{{\em JHEP} {\bfseries 07}
  (2014) 151},
\href{http://arxiv.org/abs/1308.0589}{{\ttfamily arXiv:1308.0589 [hep-th]}}.
%%CITATION = ARXIV:1308.0589;%%.

\bibitem{Lysov:2014csa}
V.~Lysov, S.~Pasterski, and A.~Strominger, ``{Low's Subleading Soft Theorem as
  a Symmetry of QED},''
  \href{http://dx.doi.org/10.1103/PhysRevLett.113.111601}{{\em Phys. Rev.
  Lett.} {\bfseries 113} no.~11, (2014) 111601},
\href{http://arxiv.org/abs/1407.3814}{{\ttfamily arXiv:1407.3814 [hep-th]}}.
%%CITATION = ARXIV:1407.3814;%%.

\bibitem{Kapec:2014zla}
D.~Kapec, V.~Lysov, and A.~Strominger, ``{Asymptotic Symmetries of Massless QED
  in Even Dimensions},''
\href{http://arxiv.org/abs/1412.2763}{{\ttfamily arXiv:1412.2763 [hep-th]}}.
%%CITATION = ARXIV:1412.2763;%%.

\bibitem{Campiglia:2015qka}
M.~Campiglia and A.~Laddha, ``{Asymptotic symmetries of QED and Weinberg's soft
  photon theorem},'' \href{http://dx.doi.org/10.1007/JHEP07(2015)115}{{\em
  JHEP} {\bfseries 07} (2015) 115},
\href{http://arxiv.org/abs/1505.05346}{{\ttfamily arXiv:1505.05346 [hep-th]}}.
%%CITATION = ARXIV:1505.05346;%%.

\bibitem{Kapec:2015ena}
D.~Kapec, M.~Pate, and A.~Strominger, ``{New Symmetries of QED},''
\href{http://arxiv.org/abs/1506.02906}{{\ttfamily arXiv:1506.02906 [hep-th]}}.
%%CITATION = ARXIV:1506.02906;%%.

\bibitem{Strominger:2015bla}
A.~Strominger, ``{Magnetic Corrections to the Soft Photon Theorem},''
  \href{http://dx.doi.org/10.1103/PhysRevLett.116.031602}{{\em Phys. Rev.
  Lett.} {\bfseries 116} no.~3, (2016) 031602},
\href{http://arxiv.org/abs/1509.00543}{{\ttfamily arXiv:1509.00543 [hep-th]}}.
%%CITATION = ARXIV:1509.00543;%%.

\bibitem{He:2015zea}
T.~He, P.~Mitra, and A.~Strominger, ``{2D Kac-Moody Symmetry of 4D Yang-Mills
  Theory},'' \href{http://dx.doi.org/10.1007/JHEP10(2016)137}{{\em JHEP}
  {\bfseries 10} (2016) 137},
\href{http://arxiv.org/abs/1503.02663}{{\ttfamily arXiv:1503.02663 [hep-th]}}.
%%CITATION = ARXIV:1503.02663;%%.

\bibitem{Dumitrescu:2015fej}
T.~T. Dumitrescu, T.~He, P.~Mitra, and A.~Strominger, ``{Infinite-Dimensional
  Fermionic Symmetry in Supersymmetric Gauge Theories},''
\href{http://arxiv.org/abs/1511.07429}{{\ttfamily arXiv:1511.07429 [hep-th]}}.
%%CITATION = ARXIV:1511.07429;%%.

\bibitem{Campiglia:2016hvg}
M.~Campiglia and A.~Laddha, ``{Subleading soft photons and large gauge
  transformations},'' \href{http://dx.doi.org/10.1007/JHEP11(2016)012}{{\em
  JHEP} {\bfseries 11} (2016) 012},
\href{http://arxiv.org/abs/1605.09677}{{\ttfamily arXiv:1605.09677 [hep-th]}}.
%%CITATION = ARXIV:1605.09677;%%.

\bibitem{Christodoulou:1993uv}
D.~Christodoulou and S.~Klainerman, {\em {The Global nonlinear stability of the
  Minkowski space}}.
\newblock {Princeton University Press, Princeton},
1993.
\newblock
%%CITATION = INSPIRE-370148;%%.

\bibitem{Hawking:1976ra}
S.~W. Hawking, ``{Breakdown of Predictability in Gravitational Collapse},''
\href{http://dx.doi.org/10.1103/PhysRevD.14.2460}{{\em Phys. Rev.} {\bfseries
  D14} (1976) 2460--2473}.
%%CITATION = PHRVA,D14,2460;%%.

\bibitem{Hawking:2015qqa}
S.~W. Hawking, ``{The Information Paradox for Black Holes},''
\newblock 2015.
\newblock
\href{http://arxiv.org/abs/1509.01147}{{\ttfamily arXiv:1509.01147 [hep-th]}}.
\newblock
%%CITATION = ARXIV:1509.01147;%%.

\bibitem{Hawking:2016msc}
S.~W. Hawking, M.~J. Perry, and A.~Strominger, ``{Soft Hair on Black Holes},''
  \href{http://dx.doi.org/10.1103/PhysRevLett.116.231301}{{\em Phys. Rev.
  Lett.} {\bfseries 116} no.~23, (2016) 231301},
\href{http://arxiv.org/abs/1601.00921}{{\ttfamily arXiv:1601.00921 [hep-th]}}.
%%CITATION = ARXIV:1601.00921;%%.

\bibitem{Donnay:2015abr}
L.~Donnay, G.~Giribet, H.~A. Gonzalez, and M.~Pino, ``{Supertranslations and
  Superrotations at the Black Hole Horizon},''
  \href{http://dx.doi.org/10.1103/PhysRevLett.116.091101}{{\em Phys. Rev.
  Lett.} {\bfseries 116} no.~9, (2016) 091101},
\href{http://arxiv.org/abs/1511.08687}{{\ttfamily arXiv:1511.08687 [hep-th]}}.
%%CITATION = ARXIV:1511.08687;%%.

\bibitem{Eling:2016xlx}
C.~Eling and Y.~Oz, ``{On the Membrane Paradigm and Spontaneous Breaking of
  Horizon BMS Symmetries},''
  \href{http://dx.doi.org/10.1007/JHEP07(2016)065}{{\em JHEP} {\bfseries 07}
  (2016) 065},
\href{http://arxiv.org/abs/1605.00183}{{\ttfamily arXiv:1605.00183 [hep-th]}}.
%%CITATION = ARXIV:1605.00183;%%.

\bibitem{Averin:2016ybl}
A.~Averin, G.~Dvali, C.~Gomez, and D.~Lust, ``{Gravitational Black Hole Hair
  from Event Horizon Supertranslations},''
  \href{http://dx.doi.org/10.1007/JHEP06(2016)088}{{\em JHEP} {\bfseries 06}
  (2016) 088},
\href{http://arxiv.org/abs/1601.03725}{{\ttfamily arXiv:1601.03725 [hep-th]}}.
%%CITATION = ARXIV:1601.03725;%%.

\bibitem{Hooft:2016itl}
G.~t. Hooft, ``{Black hole unitarity and antipodal entanglement},''
\href{http://arxiv.org/abs/1601.03447}{{\ttfamily arXiv:1601.03447 [gr-qc]}}.
%%CITATION = ARXIV:1601.03447;%%.

\bibitem{Compere:2016hzt}
G.~Comp\`{e}re and J.~Long, ``{Classical static final state of collapse with
  supertranslation memory},''
  \href{http://dx.doi.org/10.1088/0264-9381/33/19/195001}{{\em Class. Quant.
  Grav.} {\bfseries 33} no.~19, (2016) 195001},
\href{http://arxiv.org/abs/1602.05197}{{\ttfamily arXiv:1602.05197 [gr-qc]}}.
%%CITATION = ARXIV:1602.05197;%%.

\bibitem{Kapec:2016aqd}
D.~Kapec, A.-M. Raclariu, and A.~Strominger, ``{Area, Entanglement Entropy and
  Supertranslations at Null Infinity},''
\href{http://arxiv.org/abs/1603.07706}{{\ttfamily arXiv:1603.07706 [hep-th]}}.
%%CITATION = ARXIV:1603.07706;%%.

\bibitem{Hotta:2016qtv}
M.~Hotta, J.~Trevison, and K.~Yamaguchi, ``{Gravitational Memory Charges of
  Supertranslation and Superrotation on Rindler Horizons},''
  \href{http://dx.doi.org/10.1103/PhysRevD.94.083001}{{\em Phys. Rev.}
  {\bfseries D94} no.~8, (2016) 083001},
\href{http://arxiv.org/abs/1606.02443}{{\ttfamily arXiv:1606.02443 [gr-qc]}}.
%%CITATION = ARXIV:1606.02443;%%.

\bibitem{Hawking:2016sgy}
S.~W. Hawking, M.~J. Perry, and A.~Strominger, ``{Superrotation Charge and
  Supertranslation Hair on Black Holes},''
\href{http://arxiv.org/abs/1611.09175}{{\ttfamily arXiv:1611.09175 [hep-th]}}.
%%CITATION = ARXIV:1611.09175;%%.

\bibitem{Sheikh-Jabbari:2016unm}
M.~M. Sheikh-Jabbari and H.~Yavartanoo, ``{On 3d bulk geometry of Virasoro
  coadjoint orbits: orbit invariant charges and Virasoro hair on locally
  AdS$_3$ geometries},''
  \href{http://dx.doi.org/10.1140/epjc/s10052-016-4326-z}{{\em Eur. Phys. J.}
  {\bfseries C76} no.~9, (2016) 493},
\href{http://arxiv.org/abs/1603.05272}{{\ttfamily arXiv:1603.05272 [hep-th]}}.
%%CITATION = ARXIV:1603.05272;%%.

\bibitem{Blau:2016juv}
M.~Blau and M.~O'Loughlin, ``{Horizon Shells: Classical Structure at the
  Horizon of a Black Hole},''
  \href{http://dx.doi.org/10.1142/S0218271816440107}{{\em Int. J. Mod. Phys.}
  {\bfseries D25} no.~12, (2016) 1644010},
\href{http://arxiv.org/abs/1604.01181}{{\ttfamily arXiv:1604.01181 [hep-th]}}.
%%CITATION = ARXIV:1604.01181;%%.

\bibitem{Afshar:2016uax}
H.~Afshar, D.~Grumiller, and M.~M. Sheikh-Jabbari, ``{Black Hole Horizon
  Fluffs: Near Horizon Soft Hairs as Microstates of Three Dimensional Black
  Holes},''
\href{http://arxiv.org/abs/1607.00009}{{\ttfamily arXiv:1607.00009 [hep-th]}}.
%%CITATION = ARXIV:1607.00009;%%.

\bibitem{Grumiller:2016kcp}
D.~Grumiller, A.~Perez, S.~Prohazka, D.~Tempo, and R.~Troncoso, ``{Higher Spin
  Black Holes with Soft Hair},''
  \href{http://dx.doi.org/10.1007/JHEP10(2016)119}{{\em JHEP} {\bfseries 10}
  (2016) 119},
\href{http://arxiv.org/abs/1607.05360}{{\ttfamily arXiv:1607.05360 [hep-th]}}.
%%CITATION = ARXIV:1607.05360;%%.

\bibitem{Cardoso:2016ryw}
V.~Cardoso and L.~Gualtieri, ``{Testing the black hole ‘no-hair’
  hypothesis},'' \href{http://dx.doi.org/10.1088/0264-9381/33/17/174001}{{\em
  Class. Quant. Grav.} {\bfseries 33} no.~17, (2016) 174001},
\href{http://arxiv.org/abs/1607.03133}{{\ttfamily arXiv:1607.03133 [gr-qc]}}.
%%CITATION = ARXIV:1607.03133;%%.

\bibitem{Donnay:2016ejv}
L.~Donnay, G.~Giribet, H.~A. González, and M.~Pino, ``{Extended Symmetries at
  the Black Hole Horizon},''
  \href{http://dx.doi.org/10.1007/JHEP09(2016)100}{{\em JHEP} {\bfseries 09}
  (2016) 100},
\href{http://arxiv.org/abs/1607.05703}{{\ttfamily arXiv:1607.05703 [hep-th]}}.
%%CITATION = ARXIV:1607.05703;%%.

\bibitem{Sheikh-Jabbari:2016npa}
M.~M. Sheikh-Jabbari and H.~Yavartanoo, ``{Horizon Fluffs: Near Horizon Soft
  Hairs as Microstates of Generic AdS3 Black Holes},''
\href{http://arxiv.org/abs/1608.01293}{{\ttfamily arXiv:1608.01293 [hep-th]}}.
%%CITATION = ARXIV:1608.01293;%%.

\bibitem{Carlip:2016lnw}
S.~Carlip, ``{The Dynamics of Supertranslations and Superrotations in 2+1
  Dimensions},''
\href{http://arxiv.org/abs/1608.05088}{{\ttfamily arXiv:1608.05088 [gr-qc]}}.
%%CITATION = ARXIV:1608.05088;%%.

\bibitem{Cai:2016idg}
R.-G. Cai, S.-M. Ruan, and Y.-L. Zhang, ``{Horizon supertranslation and
  degenerate black hole solutions},''
  \href{http://dx.doi.org/10.1007/JHEP09(2016)163}{{\em JHEP} {\bfseries 09}
  (2016) 163},
\href{http://arxiv.org/abs/1609.01056}{{\ttfamily arXiv:1609.01056 [gr-qc]}}.
%%CITATION = ARXIV:1609.01056;%%.

\bibitem{Strominger:2016wns}
A.~Strominger and A.~Zhiboedov, ``{Superrotations and Black Hole Pair
  Creation},''
\href{http://arxiv.org/abs/1610.00639}{{\ttfamily arXiv:1610.00639 [hep-th]}}.
%%CITATION = ARXIV:1610.00639;%%.

\bibitem{Shi:2016jtn}
C.~Shi and J.~Mei, ``{Extended Symmetries at Black Hole Horizons in Generic
  Dimensions},''
\href{http://arxiv.org/abs/1611.09491}{{\ttfamily arXiv:1611.09491 [gr-qc]}}.
%%CITATION = ARXIV:1611.09491;%%.

\bibitem{Afshar:2016kjj}
H.~Afshar, D.~Grumiller, W.~Merbis, A.~Perez, D.~Tempo, and R.~Troncoso,
  ``{Soft hairy horizons in three spacetime dimensions},''
\href{http://arxiv.org/abs/1611.09783}{{\ttfamily arXiv:1611.09783 [hep-th]}}.
%%CITATION = ARXIV:1611.09783;%%.

\bibitem{Bergshoeff:2016soe}
E.~Bergshoeff, D.~Grumiller, S.~Prohazka, and J.~Rosseel, ``{Three-dimensional
  Spin-3 Theories Based on General Kinematical Algebras},''
  \href{http://dx.doi.org/10.1007/JHEP01(2017)114}{{\em JHEP} {\bfseries 01}
  (2017) 114},
\href{http://arxiv.org/abs/1612.02277}{{\ttfamily arXiv:1612.02277 [hep-th]}}.
%%CITATION = ARXIV:1612.02277;%%.

\bibitem{Hajian:2016iyp}
K.~Hajian and M.~M. Sheikh-Jabbari, ``{Redundant and Physical Black Hole
  Parameters: Is there an independent physical dilaton charge?},''
\href{http://arxiv.org/abs/1612.09279}{{\ttfamily arXiv:1612.09279 [hep-th]}}.
%%CITATION = ARXIV:1612.09279;%%.

\bibitem{Chrusciel:2012jk}
P.~T. Chrusciel, J.~L. Costa, and M.~Heusler, ``{Stationary Black Holes:
  Uniqueness and Beyond},'' \href{http://dx.doi.org/10.12942/lrr-2012-7}{{\em
  Living Rev. Rel.} {\bfseries 15} (2012) 7},
\href{http://arxiv.org/abs/1205.6112}{{\ttfamily arXiv:1205.6112 [gr-qc]}}.
%%CITATION = ARXIV:1205.6112;%%.

\bibitem{Ashtekar:1998sp}
A.~Ashtekar, C.~Beetle, and S.~Fairhurst, ``{Isolated horizons: A
  Generalization of black hole mechanics},''
  \href{http://dx.doi.org/10.1088/0264-9381/16/2/027}{{\em Class. Quant. Grav.}
  {\bfseries 16} (1999) L1--L7},
\href{http://arxiv.org/abs/gr-qc/9812065}{{\ttfamily arXiv:gr-qc/9812065
  [gr-qc]}}.
%%CITATION = GR-QC/9812065;%%.

\bibitem{Ashtekar:2001is}
A.~Ashtekar, C.~Beetle, and J.~Lewandowski, ``{Mechanics of rotating isolated
  horizons},'' \href{http://dx.doi.org/10.1103/PhysRevD.64.044016}{{\em Phys.
  Rev.} {\bfseries D64} (2001) 044016},
\href{http://arxiv.org/abs/gr-qc/0103026}{{\ttfamily arXiv:gr-qc/0103026
  [gr-qc]}}.
%%CITATION = GR-QC/0103026;%%.

\bibitem{Ashtekar:2004gp}
A.~Ashtekar, J.~Engle, T.~Pawlowski, and C.~Van Den~Broeck, ``{Multipole
  moments of isolated horizons},''
  \href{http://dx.doi.org/10.1088/0264-9381/21/11/003}{{\em Class. Quant.
  Grav.} {\bfseries 21} (2004) 2549--2570},
\href{http://arxiv.org/abs/gr-qc/0401114}{{\ttfamily arXiv:gr-qc/0401114
  [gr-qc]}}.
%%CITATION = GR-QC/0401114;%%.

\bibitem{Strominger:2014pwa}
A.~Strominger and A.~Zhiboedov, ``{Gravitational Memory, BMS Supertranslations
  and Soft Theorems},'' \href{http://dx.doi.org/10.1007/JHEP01(2016)086}{{\em
  JHEP} {\bfseries 01} (2016) 086},
\href{http://arxiv.org/abs/1411.5745}{{\ttfamily arXiv:1411.5745 [hep-th]}}.
%%CITATION = ARXIV:1411.5745;%%.

\bibitem{Newman:1961qr}
E.~Newman and R.~Penrose, ``{An Approach to gravitational radiation by a method
  of spin coefficients},''
\href{http://dx.doi.org/10.1063/1.1724257}{{\em J. Math. Phys.} {\bfseries 3}
  (1962) 566--578}.
%%CITATION = JMAPA,3,566;%%.

\bibitem{Chandrasekhar}
S.~Chandrasekhar, ``{The Newman-Penrose formalism},'' in {\em {The mathematical
  theory of black holes}}, ch.~1, pp.~40--55.
\newblock Oxford, UK, 1983.

\bibitem{Ashtekar:2001jb}
A.~Ashtekar, C.~Beetle, and J.~Lewandowski, ``{Geometry of generic isolated
  horizons},'' \href{http://dx.doi.org/10.1088/0264-9381/19/6/311}{{\em Class.
  Quant. Grav.} {\bfseries 19} (2002) 1195--1225},
\href{http://arxiv.org/abs/gr-qc/0111067}{{\ttfamily arXiv:gr-qc/0111067
  [gr-qc]}}.
%%CITATION = GR-QC/0111067;%%.

\bibitem{Luk:2011vf}
J.~Luk, ``{On the Local Existence for the Characteristic Initial Value Problem
  in General Relativity},'' \href{http://dx.doi.org/10.1093/imrn/rnr201}{{\em
  Int. Math. Res. Not.} {\bfseries 20} (2012) 4625},
\href{http://arxiv.org/abs/1107.0898}{{\ttfamily arXiv:1107.0898 [gr-qc]}}.
%%CITATION = ARXIV:1107.0898;%%.

\bibitem{laura}
L.~Donnay, {\em {Symmetries and dynamics for non-AdS backgrounds in
  three-dimensional gravity}}.
\newblock PhD thesis, Brussels U., PTM, 2016.
\newblock
\url{https://inspirehep.net/record/1497160/files/12086c_70fa83502f244cdb8b5dc3328f6e9989.pdf}.
\newblock
%%CITATION = INSPIRE-1497160;%%.

\bibitem{Barnich:2013axa}
G.~Barnich and C.~Troessaert, ``{Comments on holographic current algebras and
  asymptotically flat four dimensional spacetimes at null infinity},''
  \href{http://dx.doi.org/10.1007/JHEP11(2013)003}{{\em JHEP} {\bfseries 11}
  (2013) 003},
\href{http://arxiv.org/abs/1309.0794}{{\ttfamily arXiv:1309.0794 [hep-th]}}.
%%CITATION = ARXIV:1309.0794;%%.

\bibitem{Ashtekar:2004nd}
A.~Ashtekar, J.~Engle, and C.~Van Den~Broeck, ``{Quantum horizons and black
  hole entropy: Inclusion of distortion and rotation},''
  \href{http://dx.doi.org/10.1088/0264-9381/22/4/L02}{{\em Class. Quant. Grav.}
  {\bfseries 22} (2005) L27--L34},
\href{http://arxiv.org/abs/gr-qc/0412003}{{\ttfamily arXiv:gr-qc/0412003
  [gr-qc]}}.
%%CITATION = GR-QC/0412003;%%.

\bibitem{Ashtekar:2003hk}
A.~Ashtekar and B.~Krishnan, ``{Dynamical horizons and their properties},''
  \href{http://dx.doi.org/10.1103/PhysRevD.68.104030}{{\em Phys. Rev.}
  {\bfseries D68} (2003) 104030},
\href{http://arxiv.org/abs/gr-qc/0308033}{{\ttfamily arXiv:gr-qc/0308033
  [gr-qc]}}.
%%CITATION = GR-QC/0308033;%%.

\bibitem{Ashtekar:2005ez}
A.~Ashtekar and G.~J. Galloway, ``{Some uniqueness results for dynamical
  horizons},'' \href{http://dx.doi.org/10.4310/ATMP.2005.v9.n1.a1}{{\em Adv.
  Theor. Math. Phys.} {\bfseries 9} no.~1, (2005) 1--30},
\href{http://arxiv.org/abs/gr-qc/0503109}{{\ttfamily arXiv:gr-qc/0503109
  [gr-qc]}}.
%%CITATION = GR-QC/0503109;%%.

\bibitem{Compere:2015mza}
G.~Compère, K.~Hajian, A.~Seraj, and M.~M. Sheikh-Jabbari, ``{Extremal
  Rotating Black Holes in the Near-Horizon Limit: Phase Space and Symmetry
  Algebra},'' \href{http://dx.doi.org/10.1016/j.physletb.2015.08.027}{{\em
  Phys. Lett.} {\bfseries B749} (2015) 443--447},
\href{http://arxiv.org/abs/1503.07861}{{\ttfamily arXiv:1503.07861 [hep-th]}}.
%%CITATION = ARXIV:1503.07861;%%.

\bibitem{Compere:2015bca}
G.~Compère, K.~Hajian, A.~Seraj, and M.~M. Sheikh-Jabbari, ``{Wiggling Throat
  of Extremal Black Holes},''
  \href{http://dx.doi.org/10.1007/JHEP10(2015)093}{{\em JHEP} {\bfseries 10}
  (2015) 093},
\href{http://arxiv.org/abs/1506.07181}{{\ttfamily arXiv:1506.07181 [hep-th]}}.
%%CITATION = ARXIV:1506.07181;%%.

\bibitem{Compere:2015knw}
G.~Compère, P.~Mao, A.~Seraj, and M.~M. Sheikh-Jabbari, ``{Symplectic and
  Killing symmetries of AdS$_{3}$ gravity: holographic vs boundary
  gravitons},'' \href{http://dx.doi.org/10.1007/JHEP01(2016)080}{{\em JHEP}
  {\bfseries 01} (2016) 080},
\href{http://arxiv.org/abs/1511.06079}{{\ttfamily arXiv:1511.06079 [hep-th]}}.
%%CITATION = ARXIV:1511.06079;%%.

\bibitem{Sheikh-Jabbari:2016lzm}
M.~M. Sheikh-Jabbari, ``{Residual Diffeomorphisms and Symplectic Softs Hairs:
  The Need to Refine Strict Statement of Equivalence Principle},''
  \href{http://dx.doi.org/10.1142/S0218271816440193}{{\em Int. J. Mod. Phys.}
  {\bfseries D25} no.~12, (2016) 1644019},
\href{http://arxiv.org/abs/1603.07862}{{\ttfamily arXiv:1603.07862 [hep-th]}}.
%%CITATION = ARXIV:1603.07862;%%.

\bibitem{Seraj:2016jxi}
A.~Seraj, ``{Multipole charge conservation and implications on electromagnetic
  radiation},''
\href{http://arxiv.org/abs/1610.02870}{{\ttfamily arXiv:1610.02870 [hep-th]}}.
%%CITATION = ARXIV:1610.02870;%%.

\bibitem{Compere:2014cna}
G.~Compère, L.~Donnay, P.-H. Lambert, and W.~Schulgin, ``{Liouville theory
  beyond the cosmological horizon},''
  \href{http://dx.doi.org/10.1007/JHEP03(2015)158}{{\em JHEP} {\bfseries 03}
  (2015) 158},
\href{http://arxiv.org/abs/1411.7873}{{\ttfamily arXiv:1411.7873 [hep-th]}}.
%%CITATION = ARXIV:1411.7873;%%.

\bibitem{Conde:2016csj}
E.~Conde and P.~Mao, ``{Remarks on Asymptotic Symmetries and the Sub-leading
  Soft Photon Theorem},''
  \href{http://dx.doi.org/10.1103/PhysRevD.95.021701}{{\em Phys. Rev.}
  {\bfseries D95} (2017) 021701},
\href{http://arxiv.org/abs/1605.09731}{{\ttfamily arXiv:1605.09731 [hep-th]}}.
%%CITATION = ARXIV:1605.09731;%%.

\bibitem{Conde:2016rom}
E.~Conde and P.~Mao, ``{BMS Supertranslations and Not So Soft Gravitons},''
\href{http://arxiv.org/abs/1612.08294}{{\ttfamily arXiv:1612.08294 [hep-th]}}.
%%CITATION = ARXIV:1612.08294;%%.

\end{thebibliography}\endgroup
\bibliographystyle{utphys}

\end{document}